\documentclass[conference]{IEEEtran}
\usepackage{cite}

\ifCLASSINFOpdf
  \usepackage[pdftex]{graphicx}
  \graphicspath{{./figures/}}
  \DeclareGraphicsExtensions{.pdf,.jpeg,.png}
\else
\fi
\usepackage{amsmath}
\usepackage{algorithmic}
\usepackage{listings}
\usepackage{xcolor}

\definecolor{codegreen}{rgb}{0,0.6,0}
\definecolor{codegray}{rgb}{0.5,0.5,0.5}
\definecolor{codepurple}{rgb}{0.58,0,0.82}
\definecolor{backcolour}{rgb}{0.95,0.95,0.92}

\lstdefinestyle{abstyle}{
    backgroundcolor=\color{backcolour},   
    commentstyle=\color{codegreen},
    keywordstyle=\color{magenta},
    numberstyle=\tiny\color{codegray},
    stringstyle=\color{codepurple},
    basicstyle=\ttfamily\footnotesize,
    breakatwhitespace=false,         
    breaklines=true,                 
    captionpos=t,
    language= python,
    keepspaces=true,                 
    numbers=left,                    
    numbersep=5pt,                  
    showspaces=false,                
    showstringspaces=false,
    showtabs=false,                  
    tabsize=2,
}
\usepackage{url}

\usepackage{xcolor}
\begin{document}

\bstctlcite{IEEEexample:BSTcontrol}
%
\title{Multi-FPGA Synchronization and Data Communication for Quantum Control and Measurement
}


\author{\IEEEauthorblockN{
Yilun Xu,
Neel R. Vora,
Neelay Fruitwala,
Akel Hashim,
Abhi D. Rajagopala,
Gang Huang,
Irfan Siddiqi
}
\IEEEauthorblockA{\\Lawrence Berkeley National Laboratory, Berkeley, CA 94720, USA
\\Corresponding author: Yilun Xu (email: yilunxu@lbl.gov)}}

\maketitle

\begin{abstract}
Quantum computing has grown from novel physics experiments with a few qubits controlled by a single controller to systems with qubit counts that require multiple controllers working in concert. While a few expensive multi-board commercial solutions exist, most open-source solutions are limited to single-board radio frequency system-on-chip (RFSoC). The essential requirements for a multi-board solution are clock synchronization among multiple boards and the ability to transfer data with low latency for performing real-time feedback. In this work, we design a clock synchronization framework to distribute deterministic clock and synchronize the clock counters across multiple RFSoC boards to generate time-aligned radio frequency (RF) pulses used to control qubits. We also develop a data communication system over a fiber link to transfer quantum measurement data among multiple field-programmable gate arrays (FPGAs). This clock synchronization and data communication module has been integrated into the open-source quantum control system, QubiC, to enable the execution of quantum algorithms across multiple boards. We validate the design through bench tests and qubit experiments on a superconducting quantum processor. Qubit coherence experiments driven by two clock-synchronized boards show no measurable degradation compared to single-board operation. The data communication link, built directly on raw gigabit transceivers with a minimal custom protocol, achieves a one-way board-to-board latency of 71.2 $\pm$ 2.6 ns and enables a real-time cross-board conditional bit-flip experiment, demonstrating mid-circuit measurement and feed-forward across multiple boards well within the qubit coherence time.
\end{abstract}


%
\IEEEpeerreviewmaketitle

\section{Introduction}
\label{sec:intro}
Quantum computers, poised to tackle computational challenges that traditional systems find insurmountable in the post-Moore era, herald a transformative new era in computing and pave the way for unprecedented scientific advancements and discoveries~\cite{arute2019quantum,kim2023evidence,google2023suppressing}. 
Multiple qubit modalities, including superconducting circuits, trapped ions, and neutral atoms, are advancing rapidly in both scale and fidelity. Among them, superconducting-circuit-based quantum bits (qubits) have recently demonstrated that a quantum chip can exponentially reduce errors by scaling up with more qubits~\cite{acharya2024quantum}.

Superconducting qubits are controlled and measured using RF pulses, typically within the 4–8 GHz range. 
These pulses are usually generated by a dedicated room temperature control system~\cite{xu2021qubic,stefanazzi2022qick}.
For the quantum measurement (readout), frequency-multiplexing is employed, allowing multiple qubit readout signals to share a single physical line~\cite{heinsoo2018rapid}. 
In contrast, on the control (qubit drive) side, each qubit requires its own dedicated physical channel, requiring the digital-to-analog converter (DAC) channel count to scale linearly with the number of qubits. 
The FPGA is the core component of the control system for signal generation and information processing.
However, resources on a single FPGA board are limited. 
Therefore, as the qubit count exceeds approximately 10 qubits, multiple synchronized FPGAs must be used to control a quantum chip.

Leading companies such as Google~\cite{acharya2024quantum}, IBM~\cite{abughanem2024ibm}, have successfully controlled quantum processors with hundreds of qubits. 
However, since their control system architecture is proprietary, the research community is unable to take advantage of their latest innovations. 
The cost of commercially available control systems from Quantum Machines~\cite{quantummachines2025}, Zurich Instruments~\cite{zurichinstruments2025}, and Qblox~\cite{qblox2025} are expensive and is unaffordable for academic labs for large number of qubits.
Researchers in the quantum community are favoring RFSoC platforms as their control solution of choice, due to their combination of high density, cost-effectiveness, and powerful FPGA capabilities for quantum computing applications.
With the advanced ZCU216 evaluation board from AMD~\cite{amd2025zcu216}, the open-source quantum control community is expanding in the field of superconducting quantum control with contributions from QubiC~\cite{xu2023qubic}, QICK~\cite{stefanazzi2022qick}, Prestol~\cite{tholen2022presto}, and others.

Expanding from a single evaluation board to multiple RFSoC boards requires clock synchronization. 
A clean source clock needs to be distributed to all RFSoC boards to ensure deterministic distribution.
The most common approach is the trigger-based synchronization. 
A trigger-based synchronization typically relies on the precision of the trigger signal itself. Any jitters in the generation or distribution of the trigger signal can directly affect the synchronization accuracy. Additionally, the propagation delay of the trigger signal across different FPGAs can vary, leading to misalignment in the timing of operations.

Quantum applications necessitates all-to-all connectivity in the classical control system to perform mid-circuit measurement (MCM) and conditional feed-forward operations, which are essential for error correction, measurement-based quantum computing, and state preparation~\cite{koh2023measurement,PRXQuantum.6.010307}. 
Thus, the controller for any qubit must be capable of conditionally executing pulses based on the measurement of any other qubit. 
Consequently, measurement results (possibly processed) must be distributed across all other boards in the control system. 
This distribution must occur in real-time with low latency (on the order of hundreds of nanoseconds) to remain within the qubit coherence time, which ranges from tens to hundreds of microseconds.
For a multi-board control system, this requires data communication to support large-scale advanced quantum experiments. 
Unfortunately, there is currently no open-source solution available that supports data communication on RFSoC systems for large-scale quantum experiments.

In this work, we present a scalable clock synchronization and data communication framework, which we deploy on QubiC system~\cite{xu2023qubic}. 
We design a clock synchronization framework to distribute a deterministic clock to multiple RFSoC boards and synchronize the clock counters across these boards, ensuring that the RF pulses used to control qubits are time-aligned. 
We also develop a data communication system over a fiber link to transfer quantum measurement data among multiple FPGAs.
This solution is fully open-source and can be easily integrated as a plug-in tool for RFSoC systems.
Although we demonstrate the framework with superconducting qubits, the clock synchronization and data communication layers are independent of the qubit modality and are equally applicable to control systems for trapped ions, neutral atoms, and other platforms.
By providing this framework, we aim to enable researchers to construct large-scale quantum control systems and foster further innovation in the field. 
The open-source nature of our work invites the research community to build upon and refine it, facilitating advancements in large-scale quantum computation.

\section{Related Work}
\label{sec:related}
The integration of high-speed serial technologies into FPGA platforms has led to the development of numerous FPGA clusters~\cite{samayoa2023survey} such as Microsoft's Catapult~\cite{putnam14catapult}, Amazon F2~\cite{amazonf2}, Alibaba FaaS~\cite{alibaba}, and the Open Cloud Testbed~\cite{leeser21oct}. 
Most of these clusters focus on computation and a few exceptions connect to the analog domain. 
Notably, one of the early projects, Reconfigurable Computing Clusters (RCC), demonstrated a petascale-capable computing cluster of 64 FPGAs using gigabit transceivers~\cite{sass2007rcc} and functioned as a synchronous control system for range and scaling studies of digital wireless channel emulators with external DACs and ADCs~\cite{buscemi2013dwce}. 
Over the years, CERN has developed multiple tools to support FPGA clusters for data processing~\cite{musa2008cern,bonini2023atlas}. 
One such effort is the White Rabbit project~\cite{moreira2009whiterabbit}, which achieves sub-nanosecond synchronization using Synchronous Ethernet (SyncE) for synotization and implements IEEE 1588 Precision Time Protocol (PTP)~\cite{ieee2025ptp} for high-energy physics applications. 

For quantum computing, commercial vendors such as Zurich Instruments~\cite{wang2023zi}, Qblox~\cite{mesman2024qblox}, IQM~\cite{abdurakhimov2024iqm}, and Quantum Machines~\cite{mohseni2024qm,qmpatent} have built patented scalable systems for synchronous quantum control. 
The open-source project Qibolab~\cite{efthymiou2024qibolab} utilizes the Zurich Instrument cluster and has created a workflow to deploy quantum circuits on multiple quantum devices/hardware platforms. 
This cluster includes reference clock distribution for trigger and timestamp synchronization. 
Several non-commercial research groups have designed multi-board FPGA systems that incorporate clock synchronization; however, these systems lack integrated data communication capabilities.
Some examples of clock-synchronized systems include the Presto system~\cite{tholen2022presto}, where an extra device called \emph{Metronomo} synchronizes different Presto units with a programmable reference frequency, and the system reference clock signals align the device clock outputs. 
The ICARUS-Q~\cite{park2022icarus} system, used for superconducting qubits, synchronizes multiple RFSoC boards by distributing a single master oscillator to all boards. 
The ARTIQ systems and its Sinara hardware implements DRTIO (distributed real-time IO)~\cite{drtio} for synchronization based on White Rabbit for trapped-ion qubits~\cite{Schäfer2018}. 
For quantum error correction, where computation are intrinsically parallel, Linayage et al.~\cite{liyanage2023qecc} have designed a distributed multi-FPGA union-find decoder without synchronization. 

The work presented in this paper builds the data communication directly on raw gigabit transceivers with a minimal custom protocol for low-latency measurement-data transfer over fiber across multiple FPGA/RFSoC boards, and utilizes a minimal version of PTP for clock synchronization on ZCU216 RFSoC boards. Moreover, the current design eliminates the need for an additional backplane and is provided as an open-source solution.
\section{Clock Synchronization}
The objective of clock synchronization is to achieve precise phase alignment for all clock signals across each board and to synchronize all internal pulse-trigger time references, ensuring that all DAC signals among all devices are perfectly phase-synchronized.

Synchronization requires deterministic clock distribution to each node via matched-length cables, ensuring aligned and deterministic phase relationships. 
In this design, at the board level, a nested dual phase-locked loop (PLL) operating in zero-delay mode (ZDM) is employed~\cite{ti2025lmk04828}.  
The input and feedback frequencies of the PLLs are configured to the greatest common divisor of all relevant frequencies, with the system reference clock (SYSREF) serving as the feedback signal. 
To ensure the proper functioning of the clock synchronization system, a crucial divider reset is performed following the ZDM configuration.

Each board contains multiple tiles of DACs and ADCs that require synchronization for seamless coordination and data transfer. 
The multi-tile synchronization (MTS) process achieves this determinism by aligning local multiple frames using an external SYSREF trigger signal~\cite{amd2025mts}.
Performing MTS involves enabling all clocks and SYSREF generators, capturing the SYSREF signal, resetting the clock divider, measuring and adjusting first-in first-out (FIFO) latency, and synchronizing digital features with SYSREF dynamic update events to complete the setup.

To achieve multi-FPGA synchronization, we implement a minimal version of the PTP over a copper link, enabling bi-directional synchronization with low latency.
The minimal PTP synchronization sequence involves a series of message/pulse exchanges between the primary and secondary FPGAs.
A pulse is transmitted at time $t_1$ in the primary clock domain, and it is received by the secondary FPGA at time $t_2$ in the secondary clock domain.
Subsequently, the secondary FPGA responds with another pulse at $t_3$ time in the secondary clock domain, which is then received by the primary FPGA at time $t_4$ in the primary clock domain. 
The transit time for the sequence can be calculated as $\frac{(t_4-t_1)-(t_3-t_2)}{2}$, while the constant offset between the primary and secondary clocks can be determined as $\frac{(t_2-t_1)-(t_4-t_3)}{2}$.
To synchronize the secondary clock with the primary clock, the offset correction is applied through software, ensuring accurate alignment between the two clock domains.

\begin{figure}[!t]
\centering
\includegraphics[width=1.0\linewidth]{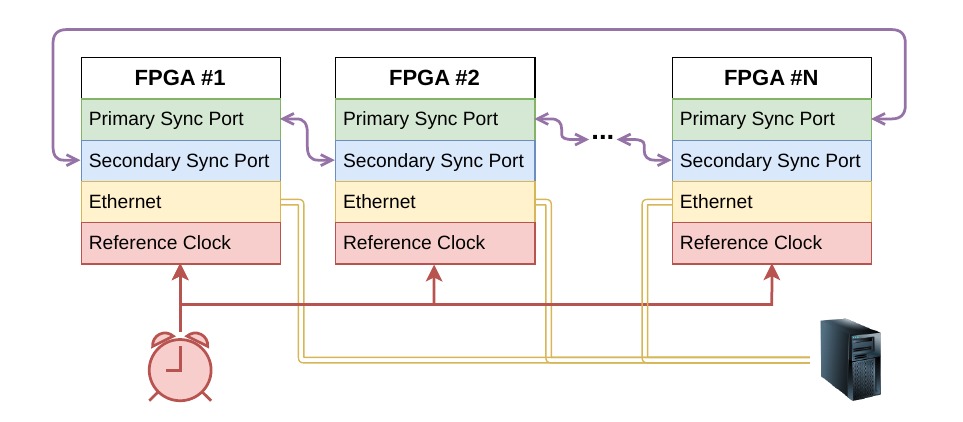}
\caption{Clock synchronization among multiple FPGA boards in a ring topology. Purple solid lines represent copper links for synchronization message/pulse transmission, red solid lines indicate matched-length cables for reference clock distribution, and yellow double lines denote communication links from the centralized job server to the FPGAs.}
\label{fig:clock_sync}
\end{figure}

We implemented clock synchronization in a ring topology for multiple FPGAs, as illustrated in Fig.~\ref{fig:clock_sync}. 
Two general-purpose input/output (GPIO) ports on each FPGA are used to connect the upstream and downstream FPGAs.
Two identical synchronization modules are instantiated in the programmable logic (PL) on each board, allowing us to use the same bitstream for all the FPGAs.
Minimal PTP is executed between each pair of neighboring FPGAs, traversing all nodes to form a ring.
Clock counter values are adjusted based on the offset calculation, and the offset correction values are propagated to the next node during the ring setup. 
Typically, offset calculation and correction are only required once during board boot-up and initialization.

We modify the software of QubiC system to integrate the clock synchronization protocol. For compiling multi-board jobs, we add a \texttt{board} attribute indicating which FPGA board to target with each compiled binary. To run a multi-board quantum program, we use a centralized job server that communicates with each individual FPGA board via remote procedure call (RPC). This job server orchestrates the minimal PTP protocol and applies corrections to synchronize the internal time reference for each board. When running a program, the job server uploads the binaries for each board, then broadcasts a \texttt{start} timestamp which triggers a synchronized start on all boards. Users can submit multi-board jobs to the job server via an RPC function call.

\section{Data Communication}
Real-time feed-forward across boards requires measurement results to be distributed with a latency that is only a small fraction of the qubit coherence time.
An earlier version of our framework employed the Aurora 64B/66B protocol~\cite{amd2025aurora} over the small form-factor pluggable (SFP) fiber-optic interfaces of the ZCU216.
While Aurora offers a general-purpose framed link, its framing, flow control, and buffering logic introduced a board-to-board latency of approximately 450 ns, consuming a significant fraction of the feed-forward budget.
We therefore re-architected the data communication to operate directly on the raw GTY gigabit transceivers with a minimum protocol tailored to quantum measurement data.

At the physical layer, the design utilizes all four SFP28 interfaces on the ZCU216.
The four GTY transceiver channels are located in two quads, and each quad shares its transceiver quad PLL, differential reference clock buffer, and clocking/reset logic between its two channels.
Each lane runs at a line rate of 25.78125 Gbps with the transceiver configured in the 64B/66B gearbox mode, providing a payload rate of 25 Gbps per lane (100 Gbps aggregate per board).
The 161.1328125 MHz transceiver reference clocks are generated by an onboard network synchronizer clock chip.

\begin{figure}[!t]
\centering
\includegraphics[width=1.0\linewidth]{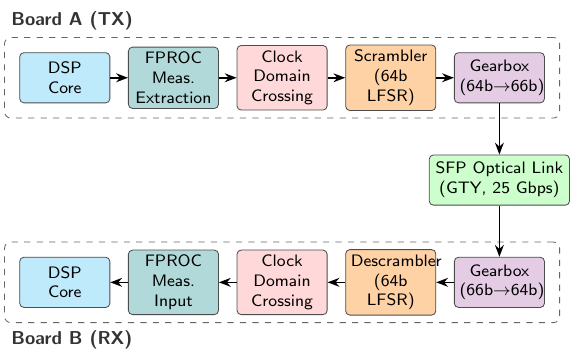}
\caption{Data communication module diagram. On the transmitting board, the discriminated measurement results extracted from the readout datapath are assembled into a 64-bit measurement word, scrambled, and streamed through the 64b-to-66b gearbox of the GTY transceiver onto a 25 Gbps SFP optical link. On the receiving board, the sync-header-based block lock aligns the incoming stream, the descrambler recovers the measurement word, and the result is injected into the function processor (FPROC) interface of the distributed processor. Each ZCU216 board provides four such lanes; for multi-FPGA data communication, this structure is duplicated on each board.}
\label{fig:data_comm}
\end{figure}

The minimum protocol strips away every latency-contributing layer of a conventional link.
The transmitter streams a 64-bit measurement word on every user clock cycle: the two-bit 64B/66B synchronization header is statically set to the data-block value, and the payload is scrambled by the standard self-synchronizing 64B/66B scrambler, $G(x) = 1 + x^{39} + x^{58}$, to guarantee the transition density required for clock recovery.
A single-cycle transmission pause is inserted periodically to compensate for the 64-to-66-bit gearbox rate difference.
On the receiver, a block synchronization state machine monitors the validity of the incoming synchronization headers and issues gearbox slip commands to shift the alignment until 64 consecutive valid headers are observed, at which point block lock is declared; the self-synchronizing descrambler then recovers the payload within 58 bits.
There are no packets, idle sequences, checksums, or handshakes on the link, and the datapath crosses between the 500 MHz control system clock domain and the transceiver user clock domain through simple two-stage register synchronizers instead of FIFO buffers.
Because the link carries a continuously refreshed state word rather than framed packets, the loss of any individual 66-bit block is self-healing, and the end-to-end latency is deterministic at the clock-cycle level.
The complete transmit and receive datapaths are illustrated in Fig.~\ref{fig:data_comm}.

The data communication module integrates directly with the distributed processor~\cite{fruitwala2023distributed} of the open-source control system through its function processor (FPROC) interface.
On the transmitting board, a measurement extraction block monitors the readout datapath of each qubit: when the readout demodulation accumulator produces a valid result, the discriminated state bit is latched into the corresponding position of a 64-bit measurement word.
The per-qubit sticky latch ensures that results are held and not overwritten, accommodating the variations in readout completion times across different qubits.
The current configuration packs one bit per qubit, supporting up to 64 qubits per word, and is extendable to multi-bit encodings for qutrit or ququart results.
The measurement word is broadcast continuously on all four lanes.
On the receiving board, the recovered 64-bit word directly drives the measurement inputs of the function processor, so each distributed processor core can read any remote measurement result, or branch on it, using the same instructions as for local measurements; multi-board feed-forward is therefore transparent to the compiled quantum program.
In a star topology, the board performing the readout acts as the root node and broadcasts its measurement word to up to four leaf nodes, whose processors conditionally execute control pulses based on the received results.
\section{Experiments}
\label{sec:exp}
Our experimental setup consists of two ZCU216 RFSoC boards connected to an external 10 MHz clock.
The clock synchronization is via copper, while the data communication is via fiber.

\subsection{Clock Synchronization}
We validated the clock synchronization functionality on two boards by measuring the alignment of the generated signals.
The individual output RF signal noise was measured using a signal source analyzer, yielding a signal root mean square (RMS) jitter of 1.8 ps (integrating from 10 Hz to 100 MHz) and 7.4 ps (integrating from 0.1 Hz to 100 MHz) at 6.50 GHz.
Additionally, we conducted a 16-hour test to validate the clock counter values and found that all boards remained synchronized with a counter offset of zero.
These results, with low jitter and zero offset over a prolong period, meet the stringent requirements of quantum computing.

\begin{figure}[!t]
\centering
\begin{minipage}{0.49\linewidth}
\centering
\includegraphics[width=\linewidth]{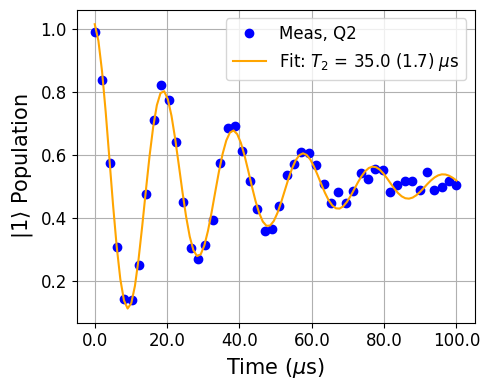}
\end{minipage}
\hfill
\begin{minipage}{0.49\linewidth}
\centering
\includegraphics[width=\linewidth]{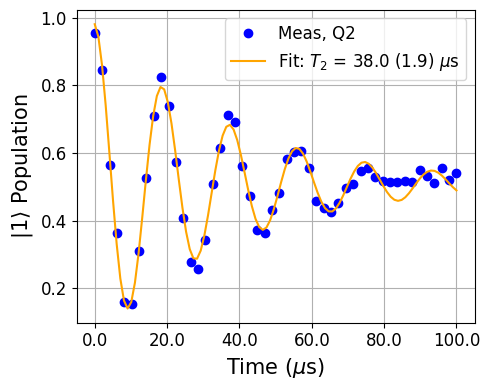}
\end{minipage}
\caption{Representative $T_2^*$ Ramsey measurements on the same qubit with a 50 kHz detuning. Left: both $X_{90}$ pulses and the readout are generated by a single board. Right: the first $X_{90}$ pulse is generated by one board and the second $X_{90}$ pulse by a second, clock-synchronized board, while the readout remains on the first board. The comparable fitted decay times confirm that the two-board timing and phase alignment does not introduce observable decoherence.}
\label{fig:ramsey}
\end{figure}

To validate the clock synchronization at the qubit level, we performed qubit coherence experiments on a superconducting eight-qubit ring chip at the LBNL Advanced Quantum Testbed (AQT)~\cite{PRXQuantum.6.010307}, with the control pulses split across two clock-synchronized ZCU216 boards.
A $T_2^*$ Ramsey experiment is a sensitive probe of both the qubit frequency and the pulse timing: any timing misalignment or phase drift between boards would manifest as a frequency offset, fringe distortion, or reduced coherence time.
We therefore ran the Ramsey sequence ($X_{90}$ -- idle delay -- $X_{90}$ -- measurement, with a 50 kHz software detuning applied through a virtual-$Z$ phase) in two configurations.
In the single-board configuration, all pulses are generated and measured by one board.
In the two-board configuration, the first $X_{90}$ pulse is driven from one board while the second $X_{90}$ pulse is driven from a second, clock-synchronized board, and the readout is still performed by the first board.
Each configuration was repeated 20 times.
The fitted coherence time is $T_2^* = 35.1 \pm 2.9$ $\mu$s for the single-board case and $T_2^* = 35.0 \pm 3.4$ $\mu$s for the two-board case (mean $\pm$ one standard deviation), which are statistically indistinguishable.
Fig.~\ref{fig:ramsey} shows a representative Ramsey fringe for each configuration: the oscillation frequency matches the programmed detuning and no phase slips are observed in the two-board case.

We repeated the comparison with a Hahn echo experiment ($T_2$ echo), in which a refocusing $\pi$ pulse is inserted between the two $X_{90}$ pulses to cancel low-frequency dephasing noise.
As in the Ramsey case, the sequence was executed either entirely from a single board, or with the pulse sequence split between two clock-synchronized boards with the final $X_{90}$ pulse driven from the second board.
Over 20 repetitions, we measured $T_2^{\mathrm{echo}} = 106.1 \pm 6.5$ $\mu$s for the single-board case and $T_2^{\mathrm{echo}} = 112.0 \pm 7.0$ $\mu$s for the two-board case.
The single-board and two-board results agree within one standard deviation for both experiments, demonstrating that the board-to-board synchronization is sufficiently precise to be invisible to time- and phase-sensitive qubit experiments.

\subsection{Data Communication}

\begin{figure}[!t]
\centering
\includegraphics[width=1.0\linewidth]{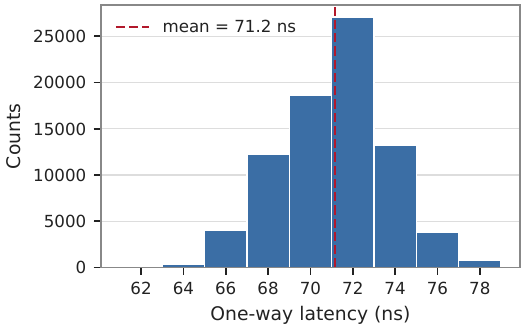}
\caption{Histogram of the measured one-way data communication latency between two ZCU216 boards over a 1 m fiber link, combining 10,000 measurements in each of the eight lane-direction combinations (80,000 total). The 2 ns binning corresponds to one cycle of the 500 MHz control system clock.}
\label{fig:latency}
\end{figure}

We first benchmarked the board-to-board communication latency.
Two ZCU216 boards were connected with 25G SFP28 optical transceivers over a 1 m fiber, and their clock counters were aligned using the copper-link synchronization described above, allowing the one-way latency to be measured directly by timestamping the transmission on one board and the reception on the other.
We collected 10,000 measurements in each direction of all four lanes (80,000 measurements in total).
As shown in Fig.~\ref{fig:latency}, the measured one-way latency is $71.2 \pm 2.6$ ns from the transmit timestamp to the receive timestamp, including the transmitter datapath, serialization, the optical link, and the receiver datapath.
This is roughly a six-fold reduction compared to the $\sim$450 ns latency of our previous Aurora-based implementation, and the sub-cycle spread confirms the deterministic behavior of the raw transceiver datapath.

We then validated the full feed-forward path with a two-qubit conditional bit-flip experiment~\cite{xu2023qubic} on the AQT eight-qubit chip, with the control and target qubits assigned to two different boards.
A control qubit (Q4), operated by the first board, is prepared on the equator of the Bloch sphere with an $X_{90}$ pulse and then measured mid-circuit.
The measurement outcome is transmitted to the second board through the data communication link.
Conditioned on receiving $|1\rangle$, the second board applies two consecutive $X_{90}$ pulses (a bit flip) to the target qubit (Q1); otherwise the target qubit is left idle.
Finally, both qubits are measured.
Ideally, the two-qubit outcomes $|00\rangle$ and $|11\rangle$ each occur with 50\% probability, while $|01\rangle$ and $|10\rangle$ are forbidden.

\begin{table}[!t]
\renewcommand{\arraystretch}{1.3}
\centering
\caption{Readout Fidelity of the Control and Target Qubits}
\label{tab:fidelity}
\begin{tabular}{l c c}
\hline
Qubit & $P(0|0)$ & $P(1|1)$\\
\hline
Control (Q4) & 0.984 & 0.975\\
Target (Q1) & 0.975 & 0.960\\
\hline
\end{tabular}

\vspace{1.5ex}

\caption{Two-Qubit Outcomes of the Cross-Board Conditional Bit-Flip Experiment (10,000 Shots)}
\label{tab:bitflip}
\begin{tabular}{l c c c c}
\hline
State & $|00\rangle$ & $|01\rangle$ & $|10\rangle$ & $|11\rangle$\\
\hline
Counts & 4155 & 288 & 379 & 3965\\
Probability & 0.4729 & 0.0328 & 0.0431 & 0.4512\\
\hline
\end{tabular}
\end{table}

The experiment was executed with 10,000 shots without any readout error mitigation.
Table~\ref{tab:fidelity} lists the independently characterized readout fidelities of the two qubits, and Table~\ref{tab:bitflip} summarizes the measured two-qubit outcomes.
The correlated outcomes $|00\rangle$ and $|11\rangle$ appear with balanced probabilities of 0.473 and 0.451 and account for 92.4\% of all shots.
The residual $|01\rangle$ and $|10\rangle$ populations are consistent with the combined readout infidelities of the two qubits, indicating that the cross-board feed-forward itself introduces no observable errors.
These results demonstrate that the clock synchronization and low-latency data communication framework supports real-time mid-circuit measurement and feed-forward across multiple boards, well within the qubit coherence time.

\section{Conclusion}
\label{sec:conc}
This work presents the first open-source solution for multi-FPGA time synchronization and data communication in the quantum domain. The synchronization ensures the qubit control pulses on multiple FPGAs are time-aligned: Ramsey and Hahn echo experiments on a superconducting quantum processor show identical coherence times whether the pulse sequence is generated by one board or split across two clock-synchronized boards. The data communication over fiber, built directly on raw gigabit transceivers with a minimal custom protocol, transfers quantum measurement data among multiple FPGAs with a one-way latency of 71.2 $\pm$ 2.6 ns. The design integrates with the QubiC system and its distributed processor to demonstrate real-time mid-circuit measurement and feed-forward across two boards through a cross-board conditional bit-flip experiment on a superconducting quantum processor. While demonstrated with superconducting qubits, the framework is agnostic to the qubit modality and can serve control systems for trapped ions, neutral atoms, and other platforms. This work is a crucial building block for implementing quantum error correction and integrating the control hardware into the data center.


\section{Acknowledgments}
\label{sec: ack}
This material is based upon work supported by the U.S. Department of Energy, Office of Science, National Quantum Information Science Research Centers, Quantum Systems Accelerator. Additional support is acknowledged from the U.S. Department of Energy, Office of Science, Advanced Scientific Computing Research Testbeds for Science program under Contract No. DE-AC02-05CH11231.

Y.X. designed and implemented the clock synchronization and data communication system. N.R.V. integrated the data communication module with the QubiC processor. N.R.V., Y.X. and A.H. performed the bench and qubit experiments. G.H., Y.X., and N.F. developed the QubiC system used in this work. A.R. provided assistance. G.H., and I.S. secured funding and provided resources. Y.X. wrote the manuscript. 

Y.X. has a financial interest in OpenQI, Inc.


\bibliographystyle{IEEEtran}
\bibliography{reference}

\end{document}